\documentstyle[aps,prb,psfig,epsfig]{revtex}            
\newcommand \be{\begin{eqnarray}}
\newcommand \ee{\end{eqnarray}}
                                               
\begin{document}
\draft
\twocolumn[\hsize\textwidth\columnwidth\hsize 
            \csname @twocolumnfalse\endcsname  
\title{Formation of three-particle clusters in hetero-junctions and
  MOSFET structures 
}
\author{E. P. Nakhmedov$^{1,2}$ and K. Morawetz$^{1}$}
\address{$^1$Max-Planck-Institute for the Physics of Complex Systems, 
Noethnitzer Str. 38, 01187 Dresden, Germany\\
$^2$Azerbaijan Academy of Sciences, Institute of Physics, H. Cavid
33, Baku, Azerbaijan}
\date{\today}
\maketitle
\begin{abstract}
A novel interaction mechanism in MOSFET structures and $GaAs/AlGaAs$
hetero--junctions between the zone electrons of the two--dimensional
(2D) gas and the charged traps on the insulator side is considered. 
By applying a canonical transformation, off--diagonal terms in the
Hamiltonian due to the trapped level
subsystem are excluded. This yields an effective three--particle 
attractive interaction as well as a pairing interaction
inside the 2D electronic band. A type of Bethe- Goldstone equation for
three particles is studied to clarify the character of the
binding
and the energy of the three-particle bound states. The results are
used to offer a possible explanation of  the Metal--Insulator transition recently observed
in MOSFET and hetero--junctions.
\end{abstract}
\pacs{71.10.-w,71.30.+h,73.20.Dx,73.30.+y}
\vskip2pc]
Significant advances in low temperature physics are particularly
connected
with the recent successes in fabricating two--dimensional (2D) electronic
structures
characterized by high mobility. 
The formation of the inversion layer of particles
on interfaces allows the study of the unusual low--temperature 
behavior of a 2D electron gas, as well as
the examination of theoretical predictions.
Particularly, 
experimental measurements at very low temperatures, $T \lesssim 2$K,
show quite nonstandard results, like the fractional quantum Hall effect
\cite{PG87}
and the Metal-Insulator transition (MIT) \cite{AKS01,pud98}, the origin
of which is not yet properly understood.
It is worthwhile to
observe that the
fractional quantum Hall effect and MIT occur only under some  special
regimes where the temperature and the impurity concentration in the
samples are very small.
On the other
hand the observation of such unusual effects only in $GaAs/AlGaAs$
hetero-junctions and $Si$--MOSFET  raises the question
whether  structural
peculiarities of these devices are responsible, 
or rather some
fundamental law is observed. 

The band structures of $GaAs/AlGaAs$
hetero-junctions and of $Si$--MOSFET are well known, and their
general features are identical despite their different
structures. The inversion of current carriers occurs at the interface of
two semiconductors (or insulator/semiconductor) with different band
gaps. These semiconductors are exclusively doped by $p$- and $n$--type
impurities in order to get high mobility in the samples. In extremely
clean samples, the doping by one type of either
acceptors or donors forms single levels in the gap of each
semiconductor. In the process of, e.g., electron inversion, the donor
centers
in $SiO_2$ of the MOSFET structure (or in, $Al_xGa_{1-x}As$
hetero-junction) become  positively charged by transferring electrons to
the 2D
electronic band. The charged donors are located within a region of at
most $\sim
200{\rm \AA}$ \cite{WG99} from the oxide--silicon interface, and their
energy level lies
above the Fermi level. Such interfacial charged states act as trap
centers for the band electrons. The density of trapped centers in,
e.g., MOSFET structures is of the order of $10^9$cm$^{-2}$ \cite{WG99},
which is considerably smaller than the typical carrier concentration of
$\sim 2\times 10^{11}$cm$^{-2}$. The Coulomb potential of charged
traps
seems to be screened imperfectly due to their low densities, and
the scattering of the band particles on these trap centers may be more
essential than the intra-band particle--particle scattering.
In this letter we study the effects of the scattering of band electrons
on
the charged traps. 

The Hamiltonian of the model can be written in the form  
$H =  H_0 + H_{int}$,
where
\be
H_0 &=& \sum_{{\bf k},\sigma} \varepsilon({\bf k}) a^+_{{\bf
    k},\sigma} a_{{\bf k},\sigma} + \sum_{\sigma} \omega_0
b^+_{o,\sigma}b_{o,\sigma} 
\label{Ho}
\ee
and
\be
H_{int}&=& \sum_{{\bf k},{\bf q};\sigma,\sigma'}
V({\bf k})[a^+_{{\bf k},\sigma}a^+_{{\bf
    -k+q},\sigma'}b_{o,\sigma'}a_{{\bf q},\sigma} \nonumber\\
&&\qquad \qquad \quad + a^+_{{\bf q},\sigma}b^+_{o,\sigma '}a_{{\bf
-k+q},\sigma '}a_{{\bf k},\sigma}].
\label{Hint}
\ee
In Eqs.(\ref{Ho}) and (\ref{Hint}), $a^+_{{\bf k},\sigma}$ and
$b^+_{o,\sigma}$ ($a_{{\bf k},\sigma}$ and
$b_{o,\sigma}$) are the creation (annihilation) operators for the
electrons in
the band and in the trapped levels, respectively. 

The trap centers are
modeled for simplicity as a dispersionless single
level. $\epsilon({\bf k})$  and $\omega_0$ in Eq.(\ref{Ho}) are the
energies of the band electrons and of the trap level,
respectively. The energy of the trap
centers is considered to be larger than the chemical
potential $\mu$ of the band electrons, $\omega_0 \gtrsim \mu$.
The first term in the Hamiltonian $H_{int}$ represents the scattering
of two band electrons via the interaction potential $V({\bf k})$
followed by the trapping of one of them by the donor level. The second
term
represents the scattering of a band electron with a trapped electron
(on a donor
level) with finally turning both of them into the band. 

For temperatures $kT< (\omega_0 - \mu )$, the trapping centers
contain a definite number of electrons at thermodynamic
equilibrium, and the considered trapping mechanism is assumed to be
essential. However, the mechanism is destroyed by increasing the gate 
potential, for then the chemical potential $\mu$ reaches the trap
level, causing the filling of all trapping centers. A temperature
increase 
also leads to the destruction of the trapping mechanism.
 
We
apply a unitary transformation \cite{SW66} with the intention to result in the
cancellation
of the off-diagonal
term, given by Eq.(\ref{Hint}). We will
show here that this unitary transformation creates a "trap mediated"
effective attraction between three electrons in the
band \cite{nak}. Notice that 
our approach to the problem is similar to the cancellation of the phonon
subsystem in superconductivity, where a canonical transformation
yields an effective attraction between electrons, see,
e.g.,\cite{Kit87}.
Let us expand a new Hamiltonian $\widetilde{H}
=e^{-iS}He^{iS}$ in power series of the operator $S =
S_1+S_2+S_3+\cdots$.
Regrouping the terms of the same order in $H_{int}$, the
conditions which
define $S_i, i=1,2,\cdots$ are obtained recursively. $S_1$ is
determined by
$
H_{int}-i[S_1,H_0]=0
$ and leads to
\begin{eqnarray}
S_1 &=& i\sum_{{\bf k},{\bf q};\sigma,\sigma '}\frac{V_{\bf
    k}}{\epsilon_{\bf k} + \epsilon_{\bf -k+q} -\epsilon_{\bf q} -
  \omega_0}
\nonumber\\
&&\times[a^+_{{\bf k},\sigma}a^+_{{\bf
    -k+q},\sigma'}b_{o,\sigma'}a_{{\bf q},\sigma} 
- a^+_{{\bf q},\sigma}b^+_{o,\sigma'}a_{{\bf -k+q},\sigma'}a_{{\bf
k},\sigma}].
\label{S}
\end{eqnarray}
The new Hamiltonian $\widetilde{H}$ can now be written in the form
$\widetilde{H} = H_0 - \frac{i}{2}[S_1,H_{int}] - i[S_2,H_0]$.
$S_2$ is obtained from the condition that the off--diagonal terms in
the
Hamiltonian be cancelled. The equation for $\widetilde {H}$ then becomes
$\widetilde{H} = H_0 - \frac{i}{2}[S_1,H_{int}]_{\rm diag}$, 
where the last term contains only diagonal elements.
Introducing Eq.\ (\ref{S}) into the expression for $\widetilde{H}$
finally yields the effective Hamiltonian
\begin{equation}
\widetilde{H}= \widetilde{H_0} + H_{e-e} + H_{\triangle}
\label{Ham}
\end{equation}
where $\widetilde{H_0}$, $H_{e-e}$ and $H_{\triangle}$ describe the
effective
one-particle Hamiltonian, the electron-electron interaction, and
the three-particle clustering, respectively.
The one-particle effective Hamiltonian $\widetilde{H_0}$ is given by
\be
\widetilde{H_0} = H_0 &-&\sum_{\sigma '}b^+_{o,\sigma '}b_{o,\sigma
  '}\sum_{{\bf q},\sigma}\varepsilon_1({\bf q})a^+_{{\bf
    q},\sigma}a_{{\bf q},\sigma}\nonumber\\
&+&\sum_{{\bf q};\sigma ,\sigma
  '} J({\bf q})b^+_{o,\sigma}b_{o,\sigma '}a^+_{{\bf q},\sigma
'}a_{{\bf
    q},\sigma},
\label{Hoeff}
\nonumber\\
\left (\matrix{\varepsilon_1({\bf q})\cr J({\bf q})} \right )&=& 
\sum_{\bf k} \frac{1}{\epsilon_{\bf
k}
  - \epsilon_{\bf -k+q}-\epsilon_{\bf q}- \omega_0}\left (\matrix{V_{\bf
    k}^2\cr V_{\bf k}V_{\bf
-k+q}}\right ).
\ee
In thermodynamical equilibrium $<b^+_{o,\sigma '}b_{o,\sigma }>
=\delta_{\sigma, \sigma '}$, and the one-particle energy is
renormalized, $\widetilde{H_0} =
\sum\limits_{{\bf
    q},\sigma}{\tilde \varepsilon (\bf q})a^+_{{\bf q},\sigma} a_{{\bf
    q},\sigma}$ with ${\tilde \varepsilon}({\bf q})= \varepsilon({\bf
q})+
  \varepsilon_1({\bf q}) + J({\bf q})$.   

The electron--electron interaction Hamiltonian $H_{e-e}$ also contains
terms with spin flipping due to the exchange scattering of 2D electrons
on 
trapped ones. In thermodynamic equilibrium, $H_{e-e}$ has the usual
form 
\be
H_{e-e}&=& \frac{1}{2}\sum_{{\bf k_1,k_2,q}; \sigma, \sigma
  '}V_{e-e}^{(eff)}({\bf k_1},{\bf k_2},{\bf q})
\nonumber\\&&\times a^+_{{\bf k_1},\sigma}
a^+_{{\bf k_2},\sigma '}a_{{\bf k_2-q},\sigma '}a_{{\bf k_1+q},\sigma},
\label{Hee}
\ee
where the effective two--particle interaction potential $ 
V_{e-e}^{(eff)}({\bf k_1},{\bf k_2},{\bf q})$ appears to be 
attractive, 
\begin{eqnarray} 
&&V_{e-e}^{(eff)}({\bf k_1},{\bf k_2},{\bf q})= 
\frac{V^2_{\bf k_1-k_2+q}}{\epsilon_{\bf
    k_1-k_2+q} + \epsilon_{\bf k_2} -\epsilon_{\bf k_1+q} -\omega_0}
\nonumber\\&&
- \frac{V_{\bf k_1}V_{\bf
    k_2-q}}{\epsilon_{\bf k_1} + \epsilon_{\bf q} -\epsilon_{\bf
    k_1+q} -\omega_0}-\frac{V_{\bf k_1}V_{\bf
    k_2-q}}{\epsilon_{\bf k_2-q} + \epsilon_{\bf q} -\epsilon_{\bf
    k_2} -\omega_0} 
\nonumber\\
&&+\frac{V_{\bf k_1}V_{\bf q} +  V_{\bf k_2-q}V_{\bf q}}{\epsilon_{\bf
q} + \epsilon_{\bf k_2-q}
  -\epsilon_{\bf k_2} -\omega_0} + \frac{V_{\bf k_1}V_{\bf q} + V_{\bf
    k_2-q}V_{\bf q}}{\epsilon_{\bf q} + \epsilon_{\bf k_1}
  -\epsilon_{\bf k_1+q} -\omega_0}   
\nonumber\\
&&+\frac{V^2_{\bf k_1-k_2+q}}{\epsilon_{\bf
    k_1-k_2+q} + \epsilon_{\bf k_2-q} -\epsilon_{\bf k_1} -\omega_0}. 
\label{Vee}
\end{eqnarray} 
Indeed, the denominator of each term in  $V_{e-e}^{(eff)}$ is
negative, owing to the fact that a donor level lies higher than the
chemical potential of band electrons.
 
The third term in the Hamiltonian (\ref{Ham}) describes an effective
three--particle scattering,
\begin{eqnarray}
&&H_{\triangle} = \frac{1}{2}
\sum\limits_{
 \stackrel{{\bf k_1},{\bf k_2},{\bf q_1},{\bf
    q_2}}
{\sigma_1,\sigma_2,\sigma_3}
}
\Big(\frac{V_{\bf k_1}V_{\bf k_2}}{\epsilon_{\bf k_1}
  + \epsilon_{\bf -k_1+q_1} - \epsilon_{\bf q_1} - \omega_0}
\nonumber\\
&&
\qquad \qquad \qquad \qquad + \frac{V_{\bf k_1}V_{\bf
k_2}}{\epsilon_{\bf k_2}
  + \epsilon_{\bf -k_2+q_2} - \epsilon_{\bf q_2} -
  \omega_0}\Big)
\nonumber\\&&\times a^+_{{\bf q_1},\sigma_1}a^+_{{\bf
k_2},\sigma_2}a^+_{{\bf
    -k_2+q_2},\sigma_3}
a_{{\bf -k_1+q_1},\sigma_3}a_{{\bf
    q_2},\sigma_2}a_{{\bf k_1},\sigma_1}.
\label{Htr}
\end{eqnarray}
Using again the condition of $\mu < \omega_0$, it is possible to see
that the strength of the three-particle interaction is negative, which
results in formation of
clusters of three electrons. 

This effective attraction among three electrons
can be understood according to the following physical argument.
The proposed mechanism of two-particle
interaction with trapping of one of the particles, 
in contrast to an
intra-band electron--electron scattering, destroys locally the
electro-neutrality of the $2D$ electron gas. The necessary
electro-neutrality
in hetero-junctions or in MOSFET's is restored by the ensuing adaption
of the height of the Schottky barrier, i.e.\, by a change in the value
of the band bending energy. However, the trapping and releasing
processes are so fast that the barrier's height cannot follow. As a
result of the trapping of band electrons, a hole appears which acts as
an attractive center for other electrons. 

The energy level of the trap centers in the above calculation is
chosen to be dispersionless for simplicity. However, even in the 
single level case, the donor center energies depend on the spatial
coordinates of the impurities due to the band bending, and therefore
become
dispersive. Including the dispersion of the trap level does not change
qualitatively our results.

We now proceed to show that a three-particle  attractive
interaction can lead to the formation of a bound state. To this end, we
consider for simplicity only the three--particle interaction, and
neglect the pairing interaction.
The Schr{\"o}dinger equation for three identical particles in 2D with
a generic
interaction potential of the form $V({\bf r_1 - r_2;r_3 - r_1;r_2 - r_3})$  
is written in the
following form after introducing the Jacobi coordinates ${\bf
  R}=\frac{1}{3}({\bf r}_1+{\bf r}_2+{\bf r}_3)$,  ${\bf r}= 
{\bf r}_3 - {\bf r}_1$ and ${\bf z} =
\frac{{\bf r}_1 + {\bf r}_3}{2} - {\bf r}_2$
\begin{eqnarray}
&&\Big\{-\frac{\hbar^2}{2m}(\frac{1}{3}\frac{\partial^2}{\partial {\bf
R}^2}
+ 2\frac{\partial^2}{\partial {\bf r}^2} +\frac{2}{3}
\frac{\partial^2}{\partial {\bf z}^2})
+ V( {\bf z} - \frac{{\bf r}}{2}; {\bf r}; {\bf z}+\frac{{\bf
r}}{2})\! \Big \} \nonumber\\
&&\times\psi({\bf R},{\bf r},{\bf z})=(\epsilon+ 3\epsilon_F) \psi({\bf R},{\bf r},{\bf z}),
\label{gold}
\end{eqnarray}
where $\epsilon$ is the three-particle excitation energy measured from
the three-particle Fermi level.
After excluding the center of mass coordinate ${\bf R}$ by expanding
$\psi({\bf R},{\bf r},{\bf z})$ in plane waves,
$\psi({\bf R},{\bf r},{\bf z})=\sum_{{\bf Q,p,q}}e^{{i\over
\hbar}({\bf Q R} +{\bf p r} +{\bf q z})}\phi({\bf p},{\bf q})$ ,
a Bethe-Goldstone-type equation, similar to the equation for Cooper
pairs
\cite{dg89}, is obtained,
\begin{eqnarray}
&&\Big(\frac{{\bf p}^2}{m}+\frac{3{\bf
    q}^2}{4m}-\epsilon -3\epsilon_F \Big)\phi ({\bf p},{\bf q})
\nonumber\\&&
+\sum_{{\bf p'},{\bf
    q'}} \tilde V({\bf p'},{\bf q'};{\bf p},{\bf q}) \, \phi ({\bf
  p'},{\bf q'}) =0.
\label{BG}
\end{eqnarray}
The interaction potential $\tilde
V({\bf p'},{\bf q'};{\bf p},{\bf q})$ is
assumed to be attractive when the energies of the three particles
(before the coordinate transformation) lie in a narrow vicinity $\hbar
\omega_0$ of the Fermi surface
$(\epsilon_F, \epsilon_F + \hbar \omega_0)$. Here $\hbar
\omega_0$ is a cut--off energy which is comparable to the order of
trap level energy measured from the Fermi level. This condition
restricts the energies, $\frac 3 4 \epsilon_F< p^2/2m < \frac 3 4
(\epsilon_F+\hbar\omega_0)$ and
$\epsilon_F< q^2/2m < \epsilon_F+\hbar\omega_0$, of the quasiparticles
obtained  after coordinates transformation.
Therefore, the  simplified attractive interaction for a system of
linear size $L$ is
$\tilde V({\bf p'},{\bf q'};{\bf p},{\bf q})= 
-\frac{V_0}{L^4}$ for
$\frac 3 4 \epsilon_F < \frac{p^2}{2m},{p'^2\over 2m} < \frac 3
4 (\epsilon_F + \hbar\omega_0)$,
$\epsilon_F < \frac{q^2}{2m},{q'^2 \over 2m} <
\epsilon_F + \hbar\omega_0 $, and
0 otherwise.

The following equation for the bound state energy of a
three--particle cluster results,
\begin{equation}
1= \frac{1}{3} V_0 N^2\int_0^{\frac 3 2 \hbar \omega_0} d\xi_1
\int_0^{\frac 3 2 \hbar
\omega_0}
d\xi_2 \frac{1}{\xi_1 + \xi_2 -\epsilon}, 
\label{E}
\end{equation}
where $N$ is the value of the density of electronic states on the
Fermi surface.  
The integration of Eq.(\ref{E})  gives an equation for
$\epsilon$,
\begin{equation}
\frac{3}{V_0 N^2\hbar \omega_0}= 3\ln
\frac{6 - 2\tilde{\epsilon}}{3 - 2\tilde{\epsilon}} - \tilde{\epsilon}
\ln \frac{-4\tilde{\epsilon}(3 -\tilde{\epsilon})}{(3 -
2\tilde{\epsilon})^2},
\label{gap}
\end{equation} 
where $\tilde{\epsilon}=\epsilon/\hbar \omega_0$ is the dimensionless
excitation energy.
The study of Eq.\ (\ref{gap}) for arbitrary negative values of
$\tilde{\epsilon}$ shows that the r.h.s.\ of this equation
is a monotonic and positive function with
a maximum value equal to $\ln{8}$ at
$\tilde{\epsilon}=0$. This imposes a lower restriction on  the
attractive potential,
$V_0 \ge \frac{1}{\ln{2} N^2\hbar \omega_0}$.
Consequently, for attractive potentials strong enough, Eq.(\ref{gap})
possesses a unique negative
solution for $-\tilde{\epsilon}\in({-\infty,0})$. This implies the
existence of three-particle bound
states.

In a weak coupling regime, when  $-\tilde{\epsilon} \ll 1$,
Eq.\ (\ref{gap}) is simplified to  
\begin{equation}
\frac{-4 \tilde{\epsilon}}{3 e} \ln \frac{-4\tilde{\epsilon}}{3 e}={4
  \over e}\left (\frac{1}{V_0 N^2\hbar \omega_0}- \ln{2}\right ),
\label{small}
\end{equation}
the solution of which does not possess a gap--like form for the
excitation energy.

In the opposite case of large negative solutions,$-\tilde{\epsilon}
\ge 3$, Eq.\ (\ref{gap})
leads immediately to
the following result for the bound
energy in the strong coupling regime, $\epsilon = -\frac{3}{4}V_0
(\hbar \omega_0 N)^2$ ,
which shows a clear perturbative non-collective behavior.

This discussion shows that molecular clustering rather than
a coherent state is realized in the system. The ground state of the 
system becomes unstable with respect to the three-particle attraction.
This seems to lead to molecular type formation with negative energy.
   
In conclusion, we want to emphasize that it is possible to understand
qualitatively the reason of MIT occurring at very low temperatures in
$GaAs/AlGaAs$ hetero-junctions and $MOSFET$
\cite{KKF94,AKS01}
in the framework of the  
formation of three-particle bound states we describe above. The
elastic 
scattering of
electrons on impurities at low temperatures, which is characterized
by a relaxation time $\tau_0$, results in the localization of all
electronic states \cite{aalr}, under the condition that $\hbar
\omega_0(n)<kT<\hbar/\tau_0$, producing an
insulating behavior for conductivity. Here $\omega_0(n)$ is the trap
level
energy measured from the Fermi 
energy,  which is a function of the 2D electron concentration $n$, or
the gate
potential. On the other hand, in a regime corresponding to $kT<\hbar
\omega_0(n)<\hbar/\tau_0$, which can be reached by varying the electron
concentration or the temperature, 
the formation of three-particles bound states results in the vanishing
of
the weak localization corrections to conductivity. This is due to the
fact that the scattering of the
three-particle clusters on the impurities does not lead to quantum
interference. Instead, the cluster's wave function accumulates an
additional
phase by rotation of the cluster in the process of scattering, while
the center of mass motion of the cluster is still extended. The
expression for the conductivity can be written as
\begin{equation}
\sigma (T)={3 e^2 n_\Delta \tau_0\over m}+{e^2 n_f\tau_0\over m}
\left (1-{\hbar \over 2\pi \epsilon_F \tau_0} \ln{{\tau_{in}(T) \over
\tau_0}}\right),
\label{sigmao}
\end{equation}
where the first and the second terms in Eq.\ (\ref{sigmao}) correspond
to the Drude and the weak localization contributions, \cite{LR85}, which
correspond to 
three-particles clusters and free band electrons with concentrations of
$n_{\Delta}$ and $n_f$ respectively. $\tau_{in}$ is the inelastic
scattering time $\tau_{in}= a T^{-p}$ where $a$ is some constant,$p
\ge 2$ and $p=2$ for probable electron--electron scattering mechanism. 
Notice that a logarithmic temperature dependence of $\sigma$ in the
metallic phase has been observed, \cite{pud98}, in high--mobility
$n-Si-MOSFET$ which is in good agreement with our
assumption. Observation of the negative low--field magnetoresistance
in the metallic phase also supports an important role of the quantum
interference effects in MIT. We neglect in Eq.(\ref{sigmao}) an
additional logarithmic quantum correction due to the 
electron--electron interactions, \cite{LR85}, which is responsible for
positive magnetoresistance  also observed in experiments, \cite{AKS01}.

\begin{figure}
\psfig{file=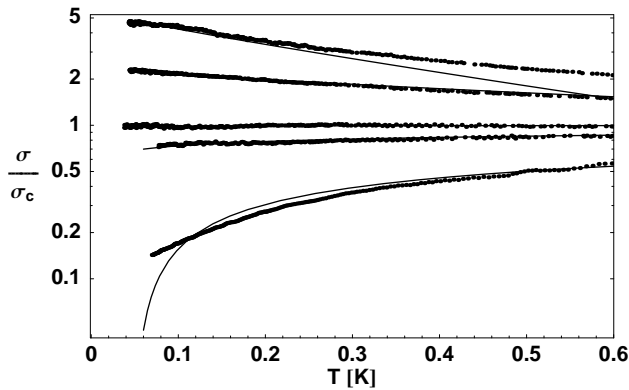,width=9cm}
\caption{The temperature dependence of the experimental (dots) conductivity scaled to the
  critical one \protect\cite{KK00}. The densities increase from below
  to above as $n=6.85, 7.17, 7.25, 7.57, 7.85$$\times 10^{10}$cm$^{-2}$.
  The lines are theoretical fits according to (\protect\ref{cond}) where
  the lower three curves are assumed to be undercritical (without $n_\Delta$) 
  with the fit
  $T_c=3.9, 3.6,  1.7$K and $\lambda=0.330, 0.078, 0.004$
  correspondingly and the above two metallic curves are
  critical (with $n_\Delta$) with  the fit $T_c=1.4, 0.8$K and $\lambda=0.48, 0.86$.\label{1}}
\end{figure}

Charge
conservation allows us to write the total concentration of particles 
as $ n= n_{\Delta} + n_f + n_t$, where $n_t$ is the concentration of
trapped electrons, which exponentially decreases with
temperature as $n_t(T)= n^o_t \exp \{-\omega(n)/kT \}$ with $n^0_t$
being the concentration of trapped impurities. Assuming that the
clustering occurs at $T=T_c(n)$, $n_{\Delta}$ can be expressed near
$T_c(n)$ as $\frac{n_{\Delta}}{n}=\frac{T_c-T}{T_c}$. We rewrite Eq.\
(\ref{sigmao}) in the form
\begin{equation}
\frac{\sigma (T)}{\sigma_0} = 1 + 2\frac{T_c-T}{T_c} -
{\hbar T \over \pi \epsilon_F \tau_0 T_c} \ln{{T^{\ast} \over T}}
\label{cond},
\end{equation}
where $\sigma_0 = e^2\tau_0 n/m$ is Drude conductivity, and $T^{\ast}=
\sqrt {a/\tau_0}$. In
Eq.(\ref{cond}) we neglected the trap level contribution due to $n_t
\ll n$. Rescaling $T$ by $T^{\ast}$ as $T/T^{\ast} \equiv \tau$ for
$\tau < 1$ and choosing 
the parameter of randomness $\lambda =
\frac{\hbar}{2\pi\epsilon_F\tau_0}\lesssim 1$ , the  unknowing
parameter $T_c/T^{\ast} \equiv \tilde{T_c}$ in 
Eq.(\ref{cond}) can be extracted by fitting $\sigma (T)/\sigma_0$ to
the experimental data. The temperature dependence of $\sigma$ is drawn in
Fig.\ref{1} for the best fit parameters $\tilde{T_c}(n)$.   

While the overall fit to MIT is satisfactorily we see deviations on
the insulating side. This seems to be due to strong localization effects in
these experiments. Effects observed in the weak localization regime, \cite{pud98},
showing a more pronounced logarithmic behaviour.  
The value of the conductivity at the critical density $G_c$ observed in the 
two experimets strongly differs from each other: in the strongly disordered case, \cite{AKS01}, $G_c \sim e^2 / h$, whereas in the weakly disordered case 
of \cite{pud98} $G_c \sim 120 e^2 /h$. Therefore, we think that 
the conductivity at the critical density $G_c$ does not show a universal 
behavior. It depends on two factors: on the impurity concentration in the $Si$ 
substrate or in the 2D electron gas 
and on the electron concentration (or on the band filling) in the inversion 
layer. By increasing the band filling in the insulator side of the MIT the 
three 
particle clusters appear which weaken the localization tendency in 
the 2D electronic system since the scattering off the cluster on the 
impurities does not lead to the quantum interference effects. At the critical 
density, the contributions coming from the clustering completely compensate 
the localization corrections and the conductivity is defined by the value of 
the residual Drude conductivity, which is temperature independent at low 
temperature.

Notice that another mechanism
 for MIT, which is
controlled by a temperature--dependent trapped--electron concentration
$n_t(T)$, has been recently proposed by
Altshuler and Maslov \cite {AM99}. As the comment and reply shows,\cite{AM991} this qualitatively
correct explanation cannot reproduce quantitative features of
the experiment. A critical discussion of different approaches
can be found in Ref.~\cite{AKS01}.

Although we have not discussed a role of effective pairing governed by
Eq.(\ref{Hee}) in the Hamiltonian, there was an attempt to
interprete the experimental data on MIT as a result of possible superconducting
ground state, \cite{Nature98}. It is well known that the effective
pairing is suppressed by the order parameter phase fluctuations in
2D systems reducing $T_{SC}$ of the superconducting transition to
zero. However, fluctuations of the order parameter modulus above $T_{SC}$
may lead to the metallic phase.

The geometry of the cluster may be either in the form of a triangle with 3/2 
and 1/2 total spin, or of a 
string--like configuration with 1/2 total spin, when two electrons with 
antiparallel spins are placed at the same point and the third electron with 
arbitrary spin is far from them. In the case of 3/2 total spin, a magnetic 
field parallel to the 
triangle area does not destroy the cluster, whereas the 
configurations with 1/2 total spin 
are destroyed due to the Zeeman effect. In both cases the magnetic 
field effects are defined by the contributions coming from the quantum 
localization corrections.

The model of three-particle clustering due to the discussed exchange type
of interaction with donor levels seems to be also a favorable candidate
for the understanding of the Fractional Quantum Hall effect. The interaction
of the band electrons with trap centers effectively leads to a formation of 
three-particle clusters, see Eq.\ (\ref{Htr}), as well as to the
superconducting fluctuations due to effective pairing interactions,
Eq.\ (\ref{Hee}). Both mechanisms decrease the ground state energy of
the system. Strong magnetic fields in the quantum Hall regime polarize
the spins of molecular clusters and a triangular geometry for the
cluster is realized due to the Pauli principle. An antisymmetric orbital wave
function of the triangular cluster will contain \emph{ab initio} the Jastrow
prefactor. The angular momentum $M=3$ of the cluster provides a natural
argument in Laughlin's theory to connect the filling factor $\nu =1/3$
of the parent states with the angular momentum $M=3$.

The authors gratefully acknowledge discussions with M. Ameduri, D.
Efremov, P. Fulde and K. Maki.

\end{document}